\documentclass[prb,twocolumn,showpacs,amsmath,nobibnotes,nofootinbib,floatfix]{revtex4}
\usepackage[latin1]{inputenc}
\usepackage{color}
\usepackage{ulem}
\usepackage{graphicx}
\usepackage{bm}
\usepackage{amssymb}
\usepackage{amsmath}

\begin{document}
\title{Spin orientation by electric current in (110) quantum wells}

\author{L.~E.~Golub}
\email{golub@coherent.ioffe.ru}
\author{E.~L.~Ivchenko}
\affiliation{Ioffe
Physical-Technical Institute of the Russian Academy of Sciences, 194021
St.~Petersburg, Russia}

\begin{abstract}
We develop a theory of spin orientation by electric current in (110)-grown semiconductor quantum wells. The controversy in the factor of two from two existed approaches is resolved by pointing out the importance of energy relaxation in this problem. The limiting cases of fast and slow energy relaxation relative to spin relaxation are considered for asymmetric (110) quantum wells. For symmetricly-doped structures the effect of spin orientation is shown to exist due to spatial fluctuations of the Rashba spin-orbit splitting. We demonstrate that the spin orientation depends strongly on the correlation length of these fluctuations as well as on the ratio of the energy and spin relaxation rates. The time-resolved kinetics of spin polarization by electric current is also governed by the correlation length being not purely exponential at slow energy relaxation. Electrical spin orientation in two-dimensional topological insulators is calculated and compared with the spin polarization induced by the magnetic field.
\end{abstract}



\maketitle

\section{Introduction}\label{sec:intro}
Creation and manipulation of electron spin by electrical means is at the heart of semiconductor spintronics.
The electron spin polarization generated by a charge current was predicted in Ref.~\onlinecite{Te_theor} and observed in bulk tellurium.~\cite{Te_exp}  As demonstrated in Refs.~\onlinecite{VaskoPrima,ALG_deform_bulk,Edelstein} the spin orientation 
due to an electrical current is also possible in semiconductor quantum well (QW) structures. 
This study was extended in Refs.~\onlinecite{ALGP,ChEM,VR,AverkievSilov,Korenev,ST,Raichev,Trushin}.
At present electrically polarized spins have been observed in
various low-dimensional materials based on GaAs, InAs, ZnSe and
GaN, see, e.g., Refs.~\onlinecite{exp_Ganichev,exp_Silov,exp_Sih,SG_EL} as well as in GaAs and InGaAs epitaxial layers.~\cite{Kato} In Ref.~\onlinecite{exp_Sih} current-induced spin polarization has been investigated in (110) AlGaAs QWs.

Phenomenologically, the average nonequilibrium free-carrier spin
${\bm s}$ is linked to a dc charge current ${\bm j}$ by a second-rank pseudotensor as follows
\begin{equation} \label{phenomen}
s_i = Q_{il} j_l\:.
\end{equation}
The mechanism most often considered for the current-induced spin polarization is a $\bm k$-linear spin-orbit splitting of electron energy spectrum described by the Hamiltonian
\begin{equation} \label{H_so}
{\cal H}_{\rm so}(\bm k) = \beta_{il} \sigma_i k_l\:,
\end{equation}
where ${\bm k}$ is the free-carrier wave vector and $\sigma_i$ ($i = x,y,z$) are the Pauli spin matrices.
The estimation for the induced spin polarization per particle reads
\begin{equation} \label{s}
s_i = - c {\Delta_{\rm so}^{(i)}(\bm k_{\rm dr}) \over  \langle E \rangle}\:.
\end{equation}
Here $c$ is a numerical coefficient,
\begin{equation}
\Delta_{\rm so}^{(i)}(\bm k_{\rm dr}) = 2 \beta_{il} k_{{\rm dr},l}\:,
\end{equation}
$\bm k_{\rm dr}=e {\bm {\mathcal E}}\tau_{\rm tr}/\hbar$ is the drift wave vector in the electric field $\bm {\mathcal E}$ controlled by the transport relaxation time $\tau_{\rm tr}$, the characteristic electron energy $\langle E \rangle$ equals to the Fermi energy $E_{\rm F}$ at low temperatures and to the thermal energy $k_{\rm B}T$ at high temperatures. Exactly it is defined as 
\begin{equation} \label{Ebar}
\langle E \rangle = \sum_{\bm k} E_k f'_0(E_k) / \sum_{\bm k} f'_0(E_k)\:,
\end{equation}
where $E_k = \hbar^2k^2/(2m)$, $m$ is the electron effective mass, 
$f_0(E_k)$ is the equilibrium Fermi function, 
and $f'_0(E_k) = df_0/dE_k$. In the following we assume $|\Delta_{\rm so}^{(i)}(\bm k_{\rm dr})|$ to be much smaller than $\langle E \rangle$.

Up to now, the theory of current-induced spin polarization has been focused on zinc-blende-lattice nanostructures grown
along the [001] crystallographic direction. In this paper we address the problem of electrical spin orientation in (110)-grown QWs. As is well-known\cite{ivch_pikus,110_Ohno,110_Karimov,110_Oestr,110_ST,110_Oestr2}, the specific property of symmetric (110)-oriented QWs is the suppression of spin-orbit splitting in the interface plane: in the coordinate frame $x \parallel [\bar{1}10]$, $y \parallel [001]$, $z \parallel [110]$ the tensor $\beta_{il}$ has only one nonzero component $\beta_{zx}$. In asymmetric (110) QWs, the terms due to the Rashba-effect with components $\beta_{xy} = - \beta_{yx}$ should be added to the Hamiltonian~\eqref{H_so}. 
However,
even in the symmetricly-doped (110)-QWs structures with two identical impurity layers separated from the QW by the spacers, the random distribution of dopant ions in the layers gives rise to a random electric field and, hence, to a random Rashba spin-orbit coupling.\cite{Glazov} Therefore, in general the spin-orbit Hamiltonian for electrons in the (110) QWs can be presented in the form
\begin{equation} \label{110ham}
{\cal H}_{\rm so}(\bm k, \bm r) = \beta\sigma_z k_x + \sigma_x \left\{ k_y, \alpha(\bm r) \right\} - \sigma_y \left\{ k_x, \alpha(\bm r) \right\}\:,
\end{equation}
where $\beta$ is a constant, the Rashba-term coefficient $\alpha(\bm r)$ is coordinate dependent, $k_l = - {\rm i} \nabla_l$ $(l=x,y)$, and the anticommutators are defined according to $\left\{ A,B \right\}=(AB+BA)/2$. We will concentrate the attention on the calculation of the spin component 
$s_z$ induced by the electric current $j_x \propto {\cal E}_x$ and described by the coefficient $Q_{zx}$ in Eq.~(\ref{phenomen}). In the next section we start with a particular case where the spatial dependence $\alpha(\bm r)$ can be ignored. In Sect.~\ref{sec:sf} we analyze the opposite limiting case of completely random spin-orbit coupling with $\alpha(\bm r)$ vanishing after averaging over the interface coordinates $x$ and $y$. In Sect.~\ref{sec:Discussion} we discuss 
the time-resolved kinetics of electrical spin orientation.

\section{asymmetric QWs}\label{sec:DP}

Let us consider an asymmetric QW without random spin-orbit coupling. It means that 
the spin-orbit Hamiltonian (\ref{110ham}) is characterized by the two constants, $\beta$ and 
$\alpha(\bm r) \equiv \alpha$.
It should be stressed that, strangely enough, two kinetic approaches are proposed to find the current-induced spin polarization, the first in Ref.~\onlinecite{ALGP} and the second in Refs.~\onlinecite{VR,Raichev}. The application of these approaches leads to the induced spin ${\bm s}$ described by the same equation (\ref{s}) but with different values of the coefficient $c$. Since this controversy has not been removed up to now, we present below a brief explanation of the existing ambiguity in values of 
the spin polarization. In the following the distribution of electrons in the wave vector ${\bm k}$ and spin is described by the spin-density matrix $\rho(\bm k)$ which can be decomposed into a sum $f(\bm k) + {\bm \sigma} \bm{\mathcal S}_{\bm k}$ that explicitly contains the scalar distribution function $f(\bm k) = {\rm Tr}\{\rho(\bm k)\}/2$ and the average spin $\bm{\mathcal S}_{\bm k} = {\rm Tr}\{{\bm \sigma} \rho(\bm k)\}/2$ of an electron with the wave vector ${\bm k}$. The average spin per particle is related to $\bm{\mathcal S}_{\bm k}$ by
\begin{equation} \label{sS}
\bm s = \sum_{\bm k} \bm{\mathcal S}_{\bm k}/N\:,
\end{equation}  
where $N$ is the two-dimensional (2D) electron density. At equilibrium, $\rho(\bm k)$ is the matrix Fermi function
\[
\rho(\bm k) = f_0[ E_k + {\cal H}_{\rm so}(\bm k)] = f_0 + {\bm \sigma}{\bm S}^0_{\bm k}\:,
\]
where $S^0_{{\bm k},i} = f_0^{\prime} \beta_{il} k_l$, 
and $f_0^{\prime} \equiv f_0^{\prime}(E_k)$.

Following Refs.~\onlinecite{ALGP,VR,AverkievSilov,Raichev} we use the coupled kinetic equations for $f(\bm k)$ and $\bm{\mathcal S}_{\bm k}$. In the first order in $|\beta_{il}|k/\langle E \rangle \ll 1$, 
can write $f(\bm k)$  in the standard form $f(\bm k) = f_0+f_1(\bm k)$, where
\begin{equation}
\label{f1}
f_1(\bm k)= -{e\hbar \tau_1\over m} f_0'\  (\bm{\mathcal E} {\bm k})\:,
\end{equation}
and reduce the equation for the electric-field induced correction ${\bm S}_{\bm k} = \bm{\mathcal S}_{\bm k} - {\bm S}^0_{\bm k}$ to 
\begin{equation} \label{kin_eq_S}
{\partial {\bm S}_{\bm k} \over \partial t} + {\bm S}_{\bm k} \times \bm \Omega_{\bm k}
= {\rm St}_p\{{\bm S}_{\bm k}\} + {\rm St}_\varepsilon\{{\bm S}_{\bm k}\} + \bm G_{\bm k}\:.
\end{equation}
Here $\bm \Omega_{\bm k}$ is the effective Larmor frequency with the components $\Omega_{\bm k,i} = 2 \beta_{il} k_l/\hbar$, ${\rm St}_p\{\bm S_{\bm k}\}$ is the elastic-scattering collision integral at zero spin-orbit coupling,
\[
{\rm St}_p\{{\bm S}_{\bm k}\} =  {2\pi\over\hbar} N_{\rm i} \sum_{\bm k'} |V_{\bm k' \bm k}|^2 \delta(E_k - E_{k'}) 
({\bm S}_{{\bm  k}'} - {\bm S}_{\bm k} )\:,
\]
$N_{\rm i}$ is the 2D density of static scatterers, $V_{\bm k' \bm k}$ is the matrix element of scattering by a single scatterer. 
The inhomogeneous term in Eq.~(\ref{kin_eq_S}) 
\begin{equation} \label{gasymm}
{\bm G}_{\bm k} = -{e\hbar\over m} {(\tau_1 f_0')' \over \tau_2} \, {\hbar\over 2} \left[\bm \Omega_{\bm k} ({\bm {\mathcal E}} \bm k) - \overline{\bm \Omega_{\bm k} ({\bm {\mathcal E}} \bm k)} \right]
\end{equation}
appears due to allowance for the linear-$\bm k$ spin-orbit splitting and the matrix expansion\cite{IvchLyandaPikus}
\begin{eqnarray} \label{deltadelta}
&&\delta[E_{k} + {\cal H}_{\rm so}({\bm k}) - E_{k'} - {\cal H}_{\rm so}({\bm k}')]
\approx \delta(E_{k} - E_{k'}) \nonumber\\ && \mbox{} \hspace{4 mm} + [{\cal H}_{\rm so}(\bm k)  - {\cal H}_{\rm so}({\bm k}')]
\frac{\partial }{\partial E_{k}} \delta(E_{k} - E_{k'})\:. 
\end{eqnarray}
Hereafter the overline means averaging over the direction of the electron wave vector.
The momentum relaxation times $\tau_1$ and $\tau_2$ are related to the scattering matrix elements by
\[
\frac{1}{\tau_n(E_k)} = \frac{2 \pi}{\hbar} N_{\rm i} \sum_{\bm k'} [1 - \cos{(n \theta)}]\ |V_{\bm k' \bm k}|^2 \delta(E_k - E_{k'})\:,
\]
where $\theta$ is the angle between $\bm k'$ and $\bm k$. The transport time is a weighted average of $\tau_1(E_k)$:
\begin{equation} \label{transport_time}
\tau_{\rm tr} = - \sum_{\bm k} \tau_1(E_k) E_k f'_0/ \sum_{\bm k} f_0\:.
\end{equation}
In this and next sections we consider the effect of a dc electric field and neglect the time derivative term in Eq.~(\ref{kin_eq_S}).

The term ${\rm St}_\varepsilon\{{\bm S}_{\bm k}\}$ describes both the energy relaxation and electron-electron collisions, it tends to equalize the degree of spin polarization of electrons with different energies. We denote the typical time of this equalization by $\tau_{\varepsilon}$ and assume $\tau_1 \ll \tau_{\varepsilon}$. Strange as it may seem, the value of the coefficient $c$ in Eq.~(\ref{s}) depends on the relation between  $\tau_{\varepsilon}$ and the spin relaxation time $\tau_s \sim (\alpha^2 k^2 \tau_1/\hbar^2)^{-1}$ in the Dyakonov-Perel' mechanism. For the limits of slow and fast energy relaxation, the value of spin polarization differs by a factor of two if $\tau_1$ is independent of the energy $E_k$, namely
\begin{equation} \label{c1214}
c = \left\{ \begin{array}{c} c_{\rm slow} = 1/4 \hspace{2 mm} {\rm if}\hspace{3 mm} \tau_{s} \ll \tau_{\varepsilon}\:,\\
c_{\rm fast}\ = 1/2 \hspace{2 mm} {\rm if}\hspace{3 mm} \tau_{s} \gg \tau_{\varepsilon}\:. \end{array}\right.
\end{equation}
It should be noted that in the both limiting cases the coefficient $c$ is independent of the times $\tau_{\varepsilon}$ and $\tau_s$, and only if they are comparable it becomes sensitive to the ratio $\tau_{\varepsilon}/\tau_s$ and lies inside the interval between 1/4 and 1/2. 

In the  limit of fast energy relaxation when $\tau_{s} \gg \tau_{\varepsilon}$, energy relaxation processes described by the operator ${\rm St}_\varepsilon\{{\bm S}_{\bm k}\}$ intensively mix the spin between electrons with different energies. Then we can take into account that, for the distribution
$f_0(E_k + {\bm \sigma} {\bm M})$ with an arbitrary fixed pseudovector ${\bm M}$, the both collision integrals in Eq.~(\ref{kin_eq_S}) vanish. This allows one to seek the solution ${\bm S}_{\bm k}$ 
in the form
\begin{equation} \label{fast}
{\bm S}^{\rm fast}_{\bm k} = N {\bm s} {f_0'(E_k)\over \sum_{\bm k}f_0'(E_k)} + \delta {\bm S}_{\bm k}\:,
\end{equation}
where ${\bm s}$ is the nonequilibrium spin per particle and the correction $\delta {\bm S}_{\bm k}$ vanishes after averaging over ${\bm k}$. Finding this correction in the first order in 
$\Omega_{\bm k} \tau_1 \ll 1$ (the collision-dominated regime), substituting it into Eq.~(\ref{kin_eq_S}) and summing over ${\bm k}$ we obtain Eq.~\eqref{s} with the coefficient $c$ given by
\begin{equation}
\label{c_fast}
	c_{\rm fast} = {\left< \tau_1 (\tau_1E_k^2)'\right>  \over 4 \tau_{\rm tr}^2 \left<E_k \right>}.
\end{equation}
Here,  similarly to Eq.~\eqref{Ebar}, the angle brackets are used to denote the functional
\begin{equation} \label{angles}
\left< \Phi(E_k) \right> = \frac{\sum_{\bm k} \Phi(E_k) f'_0(E_k)}{\sum_{\bm k} f'_0(E_k)}\:.
\end{equation}
In this notation 
the transport time $\tau_{\rm tr}$ defined by Eq.~(\ref{transport_time}) is given by
$\left<\tau_1 E_k\right>/\left<E_k\right>$.
For $\tau_1$ independent of the energy $E_k$, we find $c_{\rm fast}=1/2$, cf. Eq.~\eqref{c1214}. This value differs by the factor of 2 from the coefficient $c$ which follows from Eq.~(20) in Ref.~\onlinecite{ALGP}.

In the opposite limit of slow energy relaxation when $\tau_{s} \ll \tau_{\varepsilon}$, the term ${\rm St}_\varepsilon\{{\bm S}_{\bm k}\}$ can be omitted. In this regime, via the generation term ${\bm G}_{\bm k}$, the electric current pumps the electron spin. The spin polarization per particle is inhomogeneously distributed in energy.
The jumps of electrons in energy are slow and do not keep pace with the electrical spin pumping. 
The solution of time-independent equation (\ref{kin_eq_S}) for the induced spin polarization has the form 
\begin{equation} \label{solution}
{\bm S}_{\bm k}^{\rm slow} = -{e\hbar\over m} (\tau_1 f_0')' \, {\hbar\over 2} \bm \Omega_{\bm k} ({\bm {\mathcal E}} \bm k)\:.
\end{equation}
Here we used the following relation for elastic momentum relaxation rates of two-dimensional electrons:
\[
{\partial\over \partial E_k}\left({1\over\tau_2} \right) = {2\over E_k} \left({1\over\tau_1}-{1 \over \tau_2} \right).
\]
Substitution of the solution~\eqref{solution} into Eq.~(\ref{sS}) leads to 
\begin{equation}
\label{c_slow}
	c_{\rm slow} = {\left< \tau_1 \right>  \over 4 \tau_{\rm tr}}.
\end{equation}
For $\tau_1$ independent of the energy $E_k$, we find $c_{\rm slow}=1/4$, cf. Eq.~\eqref{c1214}.

To summarize this section, in contrast to Ref.~\onlinecite{ALGP} where the fast energy-mixing conditions were considered, a major part of theoretical activity\cite{Edelstein,VR,AverkievSilov,Raichev,Trushin} was (implicitly) focused on the theory valid if the processes of electron energy mixing are slow. The latter regime can be realized in a degenerate 2D electron gas at low temperatures while the former is important at moderate and high temperatures.

\section{Spin orientation in macroscopically-symmetric QWs}\label{sec:sf}
Now we turn to the symmetric QWs with the vanishing average, $\left< \alpha(\bm r) \right>_{\rm dis} =0$ and the correlation
function $\left< \alpha(\bm r) \alpha(\bm r')\right>_{\rm dis}$, where the angular brackets mean averaging over the 2D space.
In the following, we use the Fourier transform of the latter~\cite{Glazov}
\begin{equation} \label{C}
C_{\alpha\alpha}(q) = \int d \bm r \left< \alpha(\bm r) \alpha(\bm r')\right>_{\rm dis} \: {\rm e}^{{\rm i} \bm q (\bm r - \bm r')}\:.
\end{equation}

Let us introduce the electron mean free-path length $l = \bar{v} \tau_{\rm tr}$, where
$\bar{v} = \sqrt{2\langle E \rangle/m}$ and the correlation length of the 2D disorder $l_c$. If $l$ is small as compared with $l_c$ then the current-induced spin $s_z$ is given by Eq.~(\ref{s}) with the coefficient $c$ presented in the previous section, see Eqs.~(\ref{c_fast}),~\eqref{c_slow}. Therefore, here we will analyze the opposite limit $l_c < l$. In this case only the two diagonal components of the density matrix, $\rho_{jj} \equiv f_{j}({\bm k})$ are nonzero, where $j=\pm 1$ for the electron state with the spin component $j/2$ along the $z$ axis. The random spin-orbit interaction in the Hamiltonian Eq.~\eqref{110ham} serves as a perturbation for spin-flip scattering $({\bm k}, j) \to ({\bm k}', -j)$. The squared absolute value of the spin-flip matrix element averaged over the in-plane disorder, $W_{\bm k' \bm k} \equiv \left< |V_{\bm k', -j ; \bm k, j}|^2 \right>_{\rm dis}$, has the form
\begin{equation} \label{M_domains}
W_{\bm k' \bm k} = C_{\alpha\alpha}(\bm k' - \bm k) \left({\bm k + \bm k'\over 2} \right)^2\:.
\end{equation}
Then, the spin-flip collision integral 
reads
\begin{equation} \label{sf:Sz}
{2\pi\over\hbar} \sum_{\bm k'} W_{\bm k' \bm k} \delta(E_{{\bm k} j} - E_{{\bm k}',-j}) 
[f_{- j}(\bm k') - f_j(\bm k)] \:, 
\end{equation}
where $E_{{\bm k} j} = E_k + j \beta k_x$. 
Neglecting the spin-flip processes and taking into account the electric-field induced drift in the first order, we obtain for the distribution function 
\begin{equation} \label{fjk}
f_j(\bm k) = f_0(E_{{\bm k} j}) + f_1(\bm k - j{\bm k}_0)\:, 
\end{equation}
where $j {\bm k}_0$ is the extremum point in the subband $E_{{\bm k} j}$, the vector ${\bm k}_0$ has only a component along the $x$ axis equal to $ - m \beta/ \hbar^2$, and $f_1(\bm k)$ is given by Eq.~\eqref{f1}.
Substituting the function $f_j(\bm k)$ defined by Eq.~(\ref{fjk}) into the collision integral~(\ref{sf:Sz}) and using the identity (\ref{deltadelta}) we arrive at the kinetic equation for the field-induced spin distribution function  
\begin{eqnarray} \label{kin_eq_sf}
{\partial S_z(\bm k) \over \partial t} &+&
{2 \pi \over \hbar} \sum_{\bm k'} W_{{\bm k}' {\bm k}} \delta(E_k - E_{k'}) [{S}_z(\bm k)+{S}_z(\bm k')]
\nonumber \\ 
&=&
{\rm St}_p\{ {S}_z(\bm k)\} + {\rm St}_\varepsilon\{{S}_z(\bm k)\} + G^{({\rm dis})}_{\bm k}  \:. 
\end{eqnarray}
The second term in the left-hand side describes the spin relaxation while $G^{({\rm dis})}_{\bm k}$ gives the drift-induced pumping of the spin polarization. The latter can be written as a sum 
\begin{equation} \label{gg1g2}
G^{({\rm dis})}_{\bm k} = G^{({\rm dis},1)}_{\bm k}  + G^{({\rm dis},2)}_{\bm k} \:,
\end{equation}
where
\begin{eqnarray} \label{g1g2}
&& G^{({\rm dis},1)}_{\bm k} = {2 \pi \over \hbar} \sum_{\bm k'}  [f_1(\bm k) - f_1(\bm k')] \delta(E_k - E_{k'}) \nonumber \\
&& \mbox{} \hspace{2.5 cm} \times {\bm k}_0 ({\bm \nabla}_{\bm k'}-{\bm \nabla}_{\bm k}) W_{\bm k' \bm k}\:, \\
&& G^{({\rm dis},2)}_{\bm k} = \nonumber \\
&& {2\pi\over\hbar} (\bm k_0 \bm \nabla_{\bm k}) \sum_{\bm k'}  [f_1(\bm k')-f_1(\bm k)] \delta(E_k-E_{k'}) W_{\bm k' \bm k}\:. \nonumber
\end{eqnarray}
While deriving these equations we applied the identity
\[
F(\bm k + \bm k_0,\bm k' - \bm k_0) \approx F(\bm k ,\bm k') + \bm k_0 (\bm \nabla_{\bm k}-\bm \nabla_{\bm k'}) F(\bm k ,\bm k')\:.
\]
Again, we should consider in turn the regimes of fast and slow mixing of spin in the energy space, as compared with the spin relaxation rate $\tau_s^{-1}$, deriving equations for the coefficients $c_{\rm fast}$ and $c_{\rm slow}$.

\subsection{Fast energy relaxation, $\tau_\varepsilon \ll \tau_{\rm s}$}
\label{sec:fast}

In this regime the function ${S}_z(\bm k)$ is taken in the form of the first term in the right-hand side of 
Eq.~(\ref{fast}) with ${\bm s} \parallel z$, namely,
\begin{equation} \label{szEk} 
{S}_z({\bm k}) = N s_z {f_0'(E_k)\over \sum_{\bm k} f_0'(E_k)}\:.
\end{equation}
The spin relaxation time due to the random spin-orbit coupling is given by 
\begin{equation} \label{tausfast}
\tau^{-1}_{\rm s}  = \left< \Gamma_{\rm s}(E_k) \right>\:.
\end{equation} 
Here the angle brackets are defined in Eq.~\eqref{angles}, and $\Gamma_{\rm s}(E_k)$ is the relaxation rate of the electron spin $z$-component at the energy $E_k$: $\Gamma_{\rm s}(E_k) = (2m /\hbar^3) \overline{W_{\bm k' \bm k} }$, where the overline means a value of $W_{\bm k' \bm k}$ averaged over 
the angle between ${\bm k}$ and ${\bm k}'$ at $k' = k$. Equation~\eqref{M_domains} yields 
\begin{equation}
\label{Gamma_sf_domains}
\Gamma_{\rm s} 	= {m\over \pi \hbar^3} \Lambda_0, \quad
\Lambda_0(E_k)=\int\limits_0^{2k} dq C_{\alpha\alpha}(q) \sqrt{4k^2-q^2}.
\end{equation}

It follows from Eq.~\eqref{kin_eq_sf} that
the spin per particle is given by
\begin{equation} \label{s_fast}
s_z = {\tau_{\rm s} \over N} \sum_{\bm k} G_{\bm k}^{({\rm dis})} = {\tau_{\rm s} \over N} \sum_{\bm k} G^{({\rm dis},1)}_{\bm k}\:,
\end{equation}
because the generation term $G^{({\rm dis},2)}_{\bm k}$ is the full derivative and vanishes after summation over ${\bm k}$.
Taking into account that 
\[
({\bm \nabla}_{\bm k'}-{\bm \nabla}_{\bm k}) W_{\bm k' \bm k} = \frac{({\bm k} + {\bm k}')^2}{2} \frac{\bm q}{q}
C'_{\alpha\alpha}(q) 
\]
with $\bm q= {\bm k}' - {\bm k}$ being the scattering wave vector  and $C'_{\alpha\alpha}(q) = dC_{\alpha\alpha}(q)/dq$, we can eventually reduce the coefficient $c_{\rm fast}$
to
\begin{equation} \label{domains:fast}
c_{\rm fast} =  
{\left<  \tau_1 \Lambda_1 \right> \over 8 \tau_{\rm tr} \left< \Lambda_0 \right> }\:,
\end{equation}
where
\begin{equation}
	\label{Lambda1}
	\Lambda_1(E_k) = - \int\limits_0^{2k} dq  C'_{\alpha\alpha}(q)\ q \sqrt{4k^2-q^2}.
\end{equation}

The correlator~\eqref{C} for randomly distributed remote donors has the form~\cite{Glazov}
\begin{equation}
\label{correlator}
C_{\alpha\alpha}(q)= C_0 {\rm e}^{-q l_c}\:,
\end{equation}
where $l_c$, the length scale of variations in $\alpha$, equals to a half distance to the donor $\delta$-layers.
Then the spin relaxation rate and the coefficient $c_{\rm fast}$ reduce to
\begin{equation}
\Gamma_{\rm s}(E_k) = {C_0 m \over 2\hbar^3 l_c^2}   \xi F_1(\xi)
\end{equation}
and
\begin{equation} \label{domains:fast_exp_corr}
c_{\rm fast}= {1 \over 4 \tau_{\rm tr}} \left[ \left<  \tau_1 \right> - {\left<  \tau_1 \xi^2 F_0(\xi) \right> \over 2\left<\xi F_1(\xi)  \right>}\right]
\:,
\end{equation}
where $\xi=2kl_c = \sqrt{8m E_k}l_c/\hbar$, and 
\[
F_n(\xi)=I_n(\xi) - L_n(\xi)
\]
with
$I_n(\xi)$, $L_n(\xi)$ being the Bessel and Struve
functions.
At small correlation lengths, $\bar{k}l_c = \sqrt{2m \langle E \rangle} l_c/\hbar  \ll 1$, 
$c_{\rm fast}$ is linear in $l_c$ according to
\[
c_{\rm fast} = {l_c\over 3\pi\tau_{\rm tr}} \frac{\left< \tau_1 k^3\right>}{\left<k^2\right>}\:.
\]
In particular, if $\tau_1$ is independent of energy $E_k$, $c_{\rm fast} = k_{\rm F}l_c/(3\pi)$ for the degenerate statistics with $k_{\rm F}$ being the Fermi wave vector and 
$c_{\rm fast} = k_T l_c/(4\sqrt{\pi})$ with $k_T = \sqrt{2m k_{\rm B} T/\hbar^2}$ at high temperatures. With increasing the correlation length the coefficient $c_{\rm fast}$ monotonously decreases and saturates at 
\[
c_{\rm fast} = \frac{\left< \tau_1 k\right>}{8 \tau_{\rm tr} \left<k\right>}\:.
\]
for $\bar{k}l_c \gg 1$ (but $l_c \ll l$).
The coefficient $c_{\rm fast}$ at low temperatures is plotted in Fig.~1. 

\begin{figure}[t]
\includegraphics[width=0.9\linewidth]{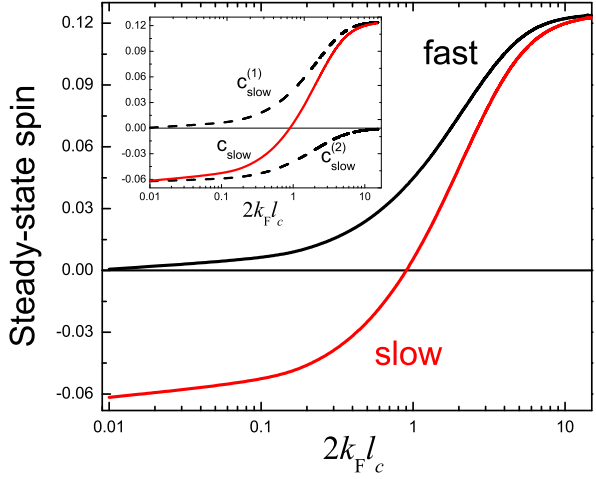}
\caption{Low-temperature spin $s_z$ in units of $-\Delta_{\rm so}^{(z)}(\bm k_{\rm dr}) / E_{\rm F}$ 
at fast  and slow  energy relaxation. 
}
\label{fig1}
\end{figure}

\subsection{Slow energy relaxation: $\tau_\varepsilon \gg \tau_{\rm s}$}

At slow energy relaxation, the spin pumped by the electric current to electrons with the energy $E_k$ is stabilized by the spin relaxation rate $\Gamma_{\rm s}(E_k)$, Eq.~\eqref{Gamma_sf_domains}. As a result the structure of spin distribution in energy strongly deviates from the quasi-equilibrium distribution~\eqref{szEk}. 
In this regime,
it is enough to average the left- and right-hand sides of Eq.~(\ref{kin_eq_sf}) over directions of ${\bm k}$, retain the axially symmetric contribution ${S}_z(E_k)$ to the function ${S}_z({\bm k})$ and obtain 
\begin{equation} \label{slowsymm}
{S}_z(E_k) =	\Gamma^{-1}_{\rm s}(E_k)\ \overline{ G^{({\rm dis})}_{\bm k}}\:.
\end{equation}
The contribution to the coefficient $c_{\rm slow}$ due to the generation term $G^{({\rm dis},1)}_{\bm k}$ can be reduced to
\begin{equation}
\label{c_slow_symm_1}
c^{(1)}_{\rm slow} =  
{1\over 8 \tau_{\rm tr}} \left< {\tau_1 \Lambda_1 \over \Lambda_0}\right> \:.
\end{equation}
The similar contribution of the second generation term, $G^{({\rm dis},2)}_{\bm k}$, has the form
\begin{equation}
\label{c_slow_symm_2}
	c^{(2)}_{\rm slow} = 
{1\over 16\tau_{\rm tr} } \left< \tau_1 \Lambda_2 E_k \left( {1\over \Lambda_0}\right)'\right>,
\end{equation}
where 
\begin{equation}
\label{Lambda_2}
\Lambda_2(E_k)= {1\over k^2}\int\limits_0^{2k} dq C_{\alpha\alpha}(q) \, q^2 \sqrt{4k^2-q^2}.
\end{equation}

For the correlator~\eqref{correlator} we obtain
\begin{eqnarray}
	&&c^{(1)}_{\rm slow} =  {1\over 4 \tau_{\rm tr} } \left< \tau_1
	\left[ 1 - {\xi F_0(\xi) \over  2F_1(\xi)} \right]
	\right>\:, \\
	&&c^{(2)}_{\rm slow} = {1\over 8 \tau_{\rm tr} } \nonumber \\ &&~ \times  \left< \tau_1
	{F_0(\xi) \left[ 3\xi F_0(\xi) - (\xi^2 + 6)F_1(\xi) + 2\xi^2/\pi \right] \over  \xi F_1^2(\xi)}
	\right>\: , \nonumber
\end{eqnarray}
where, as above, $\xi=2kl_c$.
In the limiting cases one has for the sum $c_{\rm slow} = c^{(1)}_{\rm slow}+c^{(2)}_{\rm slow}$:
\[
c_{\rm slow} = - {\left< \tau_1\right> \over 16 \tau_{\rm tr}}
\]
for small correlation lengths ($\bar{k}l_c \ll 1$), and 
\[
c_{\rm slow} = {\left< \tau_1\right> \over 8 \tau_{\rm tr}}
\]
for large correlation lengths ($\bar{k}l_c \gg 1$). 
The $l_c$-dependence of low-temperature spin $s_z$ is plotted in Fig.~1. One can see that, for the 
fast and slow energy relaxation, $s_z(l_c)$ is, respectively, a sign-preserving and a sign-changing (at $2k_{\rm F}l_c \approx 0.9$) function.

\section{Discussion}\label{sec:Discussion}

It is instructive to compare the electrically-induced spin polarization with the thermal orientation of spins in an external magnetic field. In the presence of the magnetic field ${\bm B}$, at equilibrium irrespective of the mechanisms of energy and spin relaxation, the electron spin density matrix is given by 
\begin{equation} \label{rhoB}
\rho^0_{\bm k} = f_0[E_k + {\cal H}_{\rm so}({\bm k}) + {\cal H}_{\bm B}]\:,
\end{equation}where ${\cal H}_{\bm B}$ is the Zeeman Hamiltonian $\mu_B g_{il} \sigma_i B_l/2$, $\mu_B$ is the Bohr magneton, and $g_{il}$
is the electron $g$ factor tensor. For small spin-orbit and Zeeman splittings as compared to $\langle E \rangle$ one can write instead of (\ref{rhoB})
\begin{equation} \label{rhoB2}
\rho^0_{\bm k} \approx f_0(E_k) + f'_0(E_k) [{\cal H}_{\rm so}({\bm k}) + {\cal H}_{\bm B}]\:.
\end{equation}
The magnetic-field induced spin ${\bm S}^{\bm B}_{\bm k}$ is distributed according to 
\begin{equation} \label{rhoB3}
S^{\bm B}_{\bm k,i} = \frac12 \mu_B  g_{il} B_l f'_0(E_k)\:.
\end{equation}
The spin per particle is related to the Zeeman splitting $\Delta^{(i)}_{\bm B} = \mu_B g_{il} B_l$ by
\begin{equation} \label{rhoB4}
s_i = - \frac14 \frac{\Delta_{\bm B}^{(i)}}{\langle E \rangle}\:.
\end{equation}

Contrary to the magnetic-field effect, the spin polarization induced by the electric current definitely is a nonequilibrium process and, as a result, it is dependent on mechanisms of spin relaxation as well as on the relation between spin- and energy-relaxation rates. It is interesting to notice that, even in asymmetric QWs where spin-orbit splitting disorder can be neglected, the current and magnetic-field induced polarizations differ. Indeed, for the slow energy relaxation, the coefficient
$c_{\rm slow} = 1/4$ in Eq.~(\ref{c1214}) coincides with the similar coefficient in Eq.~(\ref{rhoB4}), but the energy dependences of the spin distribution ${\bm S}^{\rm slow}_{\bm k}$ are different: according to Eqs.~(\ref{solution}) and (\ref{rhoB3}) they are proportional to $[\tau_1 f'_0(E_k)]'E_k$ and $f'_0(E_k)$, respectively. For the fast energy relaxation, the both spin distributions are proportional to $f'_0(E_k)$ but the average spins differ by the factor of 2.

In symmetric (110)-grown QWs the difference between the effects of electric current and magnetic field becomes even more striking because the former strongly depends on the correlation function of the spin-orbit disorder~(\ref{C}).

In Sects. II and III we have separately considered definitely asymmetric and
symmetric QWs, where the dispersion $\left< [\alpha(\bm r) - \alpha]^2 \right>_{\rm dis}$ is, respectively, small and large as compared to the squared
average $\alpha^2$, where $\alpha = \left< \alpha(\bm r) \right>_{\rm dis}$. If they are comparable, i.e.,
in slightly asymmetric (110) QWs, both spatially independent constant $\alpha$ and spin-orbit splitting disorder affect the current-induced spin. As a result, the electric current along the $x$ axis creates both $s_z$ and $s_y$ spin components. In this case the average spin is found from a system of two coupled kinetic equations similar to that considered in Ref.~\onlinecite{RaimSchw}. 
The contributions to generation come from both the inhomogeneous term Eq.~(\ref{gasymm}) governed by $\alpha$ and from the terms Eq.~\eqref{gg1g2} caused by spin-flip processes.  
%
Omitting details we present the result for the coefficient $c$ in Eq.~(\ref{s}). For the fast energy relaxation this coefficient is given by
\begin{equation}
c_{\rm fast} =  {c^{({\rm as})}_{\rm fast}/\tau_{\rm s}^{({\rm as})} + c^{({\rm s})}_{\rm fast}/\tau^{({\rm s})}_{\rm s} \over 1/\tau_{\rm s}^{({\rm as})} + 1/\tau^{({\rm s})}_{\rm s} }.
\end{equation}
Here $c^{({\rm as})}_{\rm fast}$ and $c^{({\rm s})}_{\rm fast}$ are the corresponding coefficients 
for asymmetric and symmetric QWs defined by Eqs.~\eqref{c_fast} and~\eqref{domains:fast}, respectively; the spin relaxation rates are given by $1/\tau_{\rm s}^{({\rm as})} = 8 \alpha^2 \tau_{\rm tr} m \left< E_k\right> / \hbar^4$ and $1/\tau^{({\rm s})}_{\rm s} = \left< \Gamma_{\rm s}(E_k)\right>$, cf. Eq.~\eqref{tausfast}. If the energy relaxation is slower than the spin relaxation we obtain 
$c_{\rm slow} = c_{\rm slow}^{({\rm as})} + c_{\rm slow}^{({\rm s})}$ with
\begin{equation}
	c_{\rm slow}^{({\rm as})} = {1\over 4 \tau_{\rm tr}} 
	\left< \tau_1 \left[E_k  {\alpha^2(2\beta^2+4\alpha^2+\eta) \over D}\right]' \right>,
\end{equation}
\begin{eqnarray}
	&& c_{\rm slow}^{({\rm s})} = {1\over 8 \tau_{\rm tr}} \\
	&& \times \left< \tau_1 \left( \Lambda_1 + {\Lambda_2 E_k\over 2}{d\over d E_k}\right) {\eta (\eta + 2\beta^2 + 7\alpha^2/2) \over \Lambda_0 \, D} \right>, \nonumber
\end{eqnarray}
\[	D(E_k) = 2(\beta^2+2\alpha^2)(\alpha^2+\eta) + \eta^2.
\]
Here $\eta(E_k)$ is the ratio between the spin relaxation rate $\Gamma_{\rm s}(E_k)$ due to spin-orbit disorder and the Dyakonov-Perel' spin relaxation rate of electrons with energy $E_k$ divided by $\alpha^2$, i.e., $\eta = \Gamma_{\rm s}\hbar^2/2\tau_1k^2$. In strongly asymmetric QWs where $\alpha^2 \gg \eta$ the equation for $c_{\rm slow}$ reduces to Eq.~\eqref{c_slow}. In the opposite limit of macroscopically-symmetric QWs, $\eta \ll \alpha^2$,  
$c_{\rm slow}$ reduces to a sum of the coefficients \eqref{c_slow_symm_1} and~\eqref{c_slow_symm_2}.

Now we turn to the discussion of the effect of energy relaxation rate on the kinetics of electrical spin orientation. In an abrupt current switching, the spin polarization builds up from zero to the steady-state value ${\bm s} \equiv {\bm s}(t \to \infty)$, see Eq.~(\ref{s}), within the spin-relaxation time $\tau_s$. If the energy relaxation is fast then we obtain for the macroscopically-symmetric (110) QWs that the spin saturation occurs according to the exponential law
\begin{equation} \label{szfastt}
s_z^{\rm fast}(t)= -{\Delta_{\rm so}^{(z)}(\bm k_{\rm dr}) \over   \left< E\right> }  \: c_{\rm fast}  \left( 1 - {\rm e}^{- {t/\tau_s}} \right)\:,
\end{equation}
where $c_{\rm fast}$ is found from Eq.~(\ref{domains:fast}) and the time $\tau_s$ is defined by Eq.~(\ref{tausfast}). The similar equation for the spin polarization $s_i^{\rm fast}(t)$ in an asymmetric QW is obtained 
from Eq.~(\ref{szfastt}) by changing $z$ to $i$ and $\tau_s$ to $\tau^{i}_s$, where $1/\tau^{i}_s$ is the principal value of the tensor\cite{Raichev}
\begin{equation}
	{1\over \tau_{s,lm}} = \left< \tau_1 \frac{\Omega_{\bm k}^2\delta_{lm} - \Omega_{\bm k,l}\Omega_{\bm k,m}}{1 + \Omega_{\bm k}^2 \tau_1^2} \right> \:.
\end{equation}
In the case of slow energy relaxation, the time variation $s_z(t)$ displays a qualitatively different behavior. Indeed, solving Eq.~\eqref{kin_eq_sf} for each fixed energy $E_k$ and then integrating the solution over $E_k$ we obtain at low temperature
\begin{eqnarray}
s_z^{\rm slow}(t)= -{\Delta_{\rm so}^{(z)}(\bm k_{\rm dr}) \over   E_{\rm F} } \hspace{2.5 cm} \\ 
\times\biggl\{ c^{(1)}_{\rm slow} \left( 1 - {\rm e}^{- {t/\tau_{\rm s}}} \right) +  c^{(2)}_{\rm slow} \left[ 1 -  \left( {t\over\tau_{\rm s}} + 1 \right) {\rm e}^{- {t/\tau_{\rm s}}} \right] \biggr\}\:, \nonumber
\end{eqnarray}
where $\tau_{\rm s} = \Gamma_{\rm s}^{-1}(E_{\rm F})$ and the relaxation rate $\Gamma_{\rm s}(E_k)$ is introduced in Eq.~(\ref{Gamma_sf_domains}). 

The calculated time-resolved kinetics of electrical spin orientation at low temperature is depicted in Fig.~2 for different values of the correlation length. One can see the difference not only in the saturation values of $s_z$ but also in the time variation at the initial stage: at slow energy relaxation, for $k_{\rm F}l_c<1$ the spin $s_z(t)$ can exhibit a nonmonotonous behavior whereas, at fast energy relaxation, only a linear increase of $s_z$ with time takes place.

\begin{figure*}[t]
\includegraphics[width=0.9\linewidth]{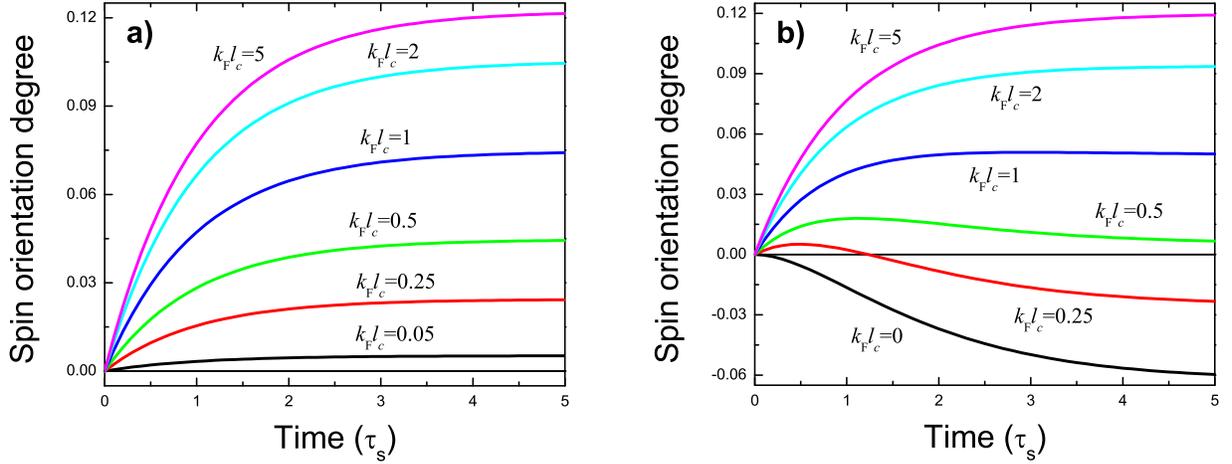}
\caption{Time-resolved kinetics of the spin $s_z$ by an abrupt current switching at low temperatures.
The spin polarization $s_z$ is presented in units of $- \Delta_{\rm so}^{(z)}(\bm k_{\rm dr}) / E_{\rm F}$ in the regimes of fast (a) and slow (b) energy relaxation. 
}
\label{fig2}
\end{figure*}

If the spin relaxation is additionally contributed by the Elliot-Yafet mechanism described in QWs by the spin-dependent scattering matrix element~\cite{AG}
\[
M_{\bm k' \bm k} = M_0(q)[\sigma_x (k_y+k'_y) - \sigma_y (k_x+k'_x)]\:,
\]
then the electrical spin polarization is given by the same expressions as in inhomogeneous QWs where the correlator $C_{\alpha\alpha}(q)$ is replaced by the sum $C_{\alpha\alpha}(q) + M_0^2(q)$. 
For another mechanism of spin-dependent scattering considered in Ref.~\onlinecite{ALGP} where
$M_{\bm k' \bm k} = M_0 \: \sigma_z (\bm k \times \bm k')_z$,
the spin $z$-component does not relax, and in this case the spin orientation by electric current is possible for the in-plane directions only.

A mechanism of electrically-induced spin orientation based on skew-scattering yields a contribution with $Q_{il} \propto \beta_{li}$,~\cite{Korenev} i.e., with the pseudotensor $\bm Q$ in Eq.~(\ref{phenomen}) being proportional to the transposed pseudotensor ${\bm \beta}$ describing the spin-orbit coupling (\ref{H_so}). One can check that, in a symmetric (110)-grown QW structure, this mechanism gives rise to no generation of the normal spin component. The reason is that, in the system of the C$_{2v}$ symmetry with the axis $z \parallel C_2 \parallel [110]$, the components $\beta_{lz}$ vanish for any $l=x,y,z$.

We finish the discussion by comparing Eq.~(\ref{s}) with the spin induced by the electric current in a topological insulator. Similarly to Ref.~\onlinecite{Hosur} we take the electron effective Hamiltonian in the form 
$${\cal H} = \hbar v_0 (\sigma_x k_y - \sigma_y k_x),
$$ 
where $v_0$ is a constant positive parameter.
The electron energy spectrum is given by two branches $E_{\pm,{\bm k}} = \pm \hbar v_0 k$ and, for typical values of $k$, the splitting between the branches, $2 \hbar v_0 k$, exceeds by far the uncertainty $\hbar/\tau_1$. Furthermore, we assume the degenerate statistics with the Fermi energy lying in the conduction band, $E_{\rm F} > 0$. The final result for the average spin per conduction-band particle reads
\[
s_x = \frac{k_{{\rm dr},y}}{2 k_{\rm F}}\:,
\qquad s_y = - \frac{k_{{\rm dr},x}}{2 k_{\rm F}}\:,
\]
or, in a more general form,
\begin{equation} \label{topological}
s_i = \frac14 \frac{ \Delta^{(i)} }{ E_{\rm F}}\:, 
\end{equation}
where, as before, $k_{\rm F} = E_{\rm F}/(\hbar v_0)$, ${\bm k}_{\rm dr} = e \tau_{\rm tr} \bm{\mathcal E}/\hbar$ and $\Delta^{(i)} = 2 \beta_{il} k_{{\rm dr},l}$, $\beta_{xy} = - \beta_{yx} = \hbar v_0$. An external in-plane magnetic field ${\bm B}$ changes the electron energy $E_{+,{\bm k}}$ by
$g \mu_B (k_y B_x - k_x B_y)/(2k)$ and the average spin is given by Eq.~(\ref{topological}), where $g$ is the electron $g$ factor and $\Delta^{(i)} = g \mu_B B_i$. Therefore, in the particular case of a topological insulator with well-resolved states the coefficient (1/4) relating $s_i$ with the ratio $\Delta^{(i)}/ E_{\rm F}$ is independent of the relaxation mechanisms and coincides with the analogous coefficient for the magnetic-field induced spin.

\section{Conclusion}\label{sec:Conclusion}

The theory of current-induced generation of electron spin polarization has been extended on quantum-well structures grown along the axis [110] from zinc-blende semiconductors. It has been shown that a value of the steady-state average spin depends on the relation between energy and spin relaxation rates. In the regime of slow energy relaxation, the spin orientation by the electric current is formed for each electron energy $E$ independently: the spin induced in the subsystem of electrons with the energy $E_k \approx E$ is stabilized by the fast energy-conserving spin relaxation, the slow energy relaxation intermixes different subsystems without affecting the stabilized spin distribution in energy. In the opposite regime of fast energy relaxation, the spins slowly generated at particular energies are rapidly intermixed in the energy space to form a quasi-equilibrium spin density matrix unambiguously determined by the average spin polarization. The analysis of these two regimes removes a controversy between the existing approaches to calculate the current-induced spin in (001)-grown QWs. 

In a symmetric (110) quantum well, the Dyakonov-Perel' mechanism cannot participate in the spin generation by electric current. It has been shown that the spin orientation can be mediated by spin-orbit splitting disorder due to a random electric field created by dopant ions located in the side $n$-doping layers. We have calculated the dependence of both the steady-state and the time resolved spin polarization on the correlation length of the disorder and showed a striking difference in the spin behaviour for the cases of fast and slow energy relaxation. 

The spin-orbit disorder can also play the dominant role in electric spin orientation in quantum wires with the spin-orbit Hamiltonian\cite{GlazovZhETP} ${\cal H}_{\rm so} = \hbar {\bm \Omega}_{k_z}{\bm \sigma}/2$, where ${\bm \Omega}_{k_z} = {\bm \lambda}k_z$, ${\bm \lambda}$ is a constant vector and $k_z$ is the wave vector of electron free motion along the wire principal axis $z$. In this case the spin oriented along ${\bm \lambda}$ is insensitive to the Dyakonov-Perel' spin relaxation mechanism and the disorder-induced spin relaxation becomes important.\cite{Glazov_wires} The developed theory can be applied as well to SiGe/Si quantum well structures with antiphase microscopic domains containing an odd number of atomic planes and shifted with respect to each other by one monoatomic layer.\cite{GeSi}

\acknowledgments{We thank M.M.~Glazov and S.A.~Tarasenko for discussions.
The work was supported by RFBR and President grant for young scientists.}


\begin{thebibliography}{24}

\bibitem{Te_theor} E.L. Ivchenko and G.E. Pikus, JETP Lett. \textbf{27}, 604 (1978) [Pisma Zh. Eksp. Teor. Fiz. \textbf{27}, 640 (1978)].
\bibitem{Te_exp} L.E. Vorob'ev, E.L. Ivchenko, G.E. Pikus, I.I. Farbstein, V.A. Shalygin, and A.V. Shturbin, JETP Lett. \textbf{29}, 441 (1979) [Pisma Zh. Eksp. Teor. Fiz. \textbf{29}, 485 (1979)].
\bibitem{VaskoPrima} F.T. Vas'ko and N.A. Prima, Sov. Phys. Solid State \textbf{21}, 994 (1979) [Fiz. Tverd. Tela \textbf{21}, 1734 (1979)].
\bibitem{ALG_deform_bulk} A.G. Aronov and Yu.B. Lyanda-Geller, JETP Lett. \textbf{50}, 431 (1989) [Pis'ma Zh. Eksp. Teor. Fiz. \textbf{50}, 398 (1989)].
\bibitem{Edelstein} V.M. Edelstein, Sol. State Commun. \textbf{73}, 233 (1990).
\bibitem{ALGP} A.G. Aronov, Yu.B. Lyanda-Geller and G.E. Pikus, JETP \textbf{100}, 973 (1991) [Sov. Phys. JETP \textbf{73}, 537 (1991)].
\bibitem{ChEM} A.V. Chaplik, M.V. Entin, and L.I. Magarill, Physica E \textbf{13}, 744 (2002).
\bibitem{VR} \textit{Quantum Kinetic Theory and Applications},  F. T. Vasko and O. E. Raichev (Springer, New York, 2005).
\bibitem{AverkievSilov} N.S. Averkiev and A.Yu. Silov, Semicond. \textbf{39}, 1323 (2005) [Fiz. Techn. Poluprov. \textbf{39}, 1370 (2005)].
\bibitem{Korenev} V.L. Korenev, Phys. Rev. B \textbf{74}, 041308 (2006).
\bibitem{ST} S.A. Tarasenko, JETP Lett. \textbf{84}, 199 (2006) [Pis'ma Zh. Eksp. Teor. Fiz. \textbf{84}, 233 (2006)]. 
\bibitem{Raichev} O. E. Raichev, Phys. Rev. B \textbf{75}, 205340 (2007).
\bibitem{Trushin} M. Trushin and J. Schliemann, Phys. Rev. B \textbf{75}, 155323 (2007).
\bibitem{exp_Ganichev} S.D. Ganichev, S.N. Danilov, Petra Schneider, V.V. Bel'kov, L.E. Golub,
W. Wegscheider, D.Weiss, and W. Prettl, cond-mat/0403641 (2004), see also J. Magn. Magn. Mater. \textbf{300}, 127 (2006).
\bibitem{exp_Silov} A.Yu. Silov, P.A. Blajnov, J.H. Wolter, R. Hey, K.H. Ploog, and N.S. Averkiev, Appl. Phys. Lett. \textbf{85}, 5929 (2004).
\bibitem{exp_Sih} V. Sih, R.C. Myers, Y.K. Kato, W.H. Lau, A.C. Gossard, and D.D. Awschalom, Nature Phys. \textbf{1}, 31 (2005).
\bibitem{SG_EL} 
S.D. Ganichev and E.L. Ivchenko 
in \textit{Spin Physics in Semiconductors}, ed. M.I. Dyakonov (Springer, 2008).
\bibitem{Kato} Y. Kato, R.C. Myers, A.C. Gossard, and D.D. Awschalom, Phys. Rev. Lett. \textbf{93}, 176601 (2004). 
\bibitem{ivch_pikus} {\it Superlattices
and Other Heterostructures. Symmetry and Optical Phenomena},  E.L. Ivchenko and G.E. Pikus, Springer Series in Solid State Sciences, Vol. 110 (Springer-Verlag,
Heidelberg, 2nd ed., 1997).
\bibitem{110_Ohno}Y. Ohno, R. Terauchi, T. Adachi, F. Matsukura, and H. Ohno, Phys. Rev. Lett. \textbf{83}, 4196 (1999).
\bibitem{110_Karimov} O. Z. Karimov, G. H. John, R. T. Harley, W. H. Lau, M. E. Flatte, M. Henini, and R. Airey, Phys. Rev. Lett. \textbf{91}, 246601 (2003) 
\bibitem{110_Oestr} S. D\"ohrmann, D. H\"agele, J. Rudolph, M. Bichler, D. Schuh, and M. Oestreich,  Phys. Rev. Lett. \textbf{93}, 147405 (2004).
\bibitem{110_ST} S.A. Tarasenko, Phys. Rev. B \textbf{80}, 165317 (2009).
\bibitem{110_Oestr2} J. H\"ubner, S. Kunz, S. Oertel, D. Schuh, M. Pochwa\l{}a, H. T. Duc, J. F\"orstner, T. Meier, and M. Oestreich,
Phys. Rev. B {\bf 84}, 041301 (2011).
\bibitem{Glazov} M. M. Glazov, E. Ya. Sherman, and V. K. Dugaev, Physica E \textbf{42}, 2157 (2010). 
\bibitem{IvchLyandaPikus} E.L. Ivchenko, Yu.B. Lyanda-Geller, and G.E. Pikus, Zh. Eksp. Teor. Fiz. {\bf 98}, 989 (1990)
[Sov. Phys. JETP {\bf 71}, 550 (1990)].
\bibitem{RaimSchw} R. Raimondi and P. Schwab, EPL \textbf{87}, 37008 (2009). 
\bibitem{AG} N. S. Averkiev and L. E. Golub, Semicond. Sci. Technol. \textbf{23}, 114002 (2008).
\bibitem{Hosur} P. Hosur, Phys. Rev. B {\bf 83}, 035309 (2011).
\bibitem{GlazovZhETP} M. M. Glazov and E. L. Ivchenko, JETP {\bf 99}, 1279 (2004).
\bibitem{Glazov_wires} M. M. Glazov and E. Ya. Sherman, arXiv:1103.6040 (2011).
\bibitem{GeSi} L. E. Golub and E. L. Ivchenko, Phys. Rev. B \textbf{69}, 115333 (2004).
\end{thebibliography}
\end{document}